\documentclass[a4paper,fleqn,usenatbib]{mnras}

\usepackage[T1]{fontenc}
\usepackage{ae,aecompl}

\usepackage[normalem]{ulem}
\usepackage{amsmath}
\usepackage{graphicx} 
\usepackage{lscape}
\usepackage{indentfirst}
\usepackage{enumitem}
\usepackage{xspace}

\usepackage{amssymb}
\usepackage[dvipsnames]{xcolor}
\usepackage{xcolor}
\usepackage{url}
\usepackage[flushleft]{threeparttable}

\newcommand {\bc}{\begin {center}}
\newcommand {\ec}{\end {center}}
\newcommand {\be}{\begin {equation}}
\newcommand {\ee}{\end {equation}}
\newcommand {\beq}{\begin {eqnarray}}
\newcommand {\eeq}{\end {eqnarray}}
\newcommand {\comment}[1]{}
\newcommand{\lambdabar}{{\mkern0.75mu\mathchar '26\mkern -9.75mu\lambda}}

\renewcommand{\d}{{\rm d}}

\newcommand {\ergs}{{\rm erg\ \rm s^{-1}}}

\title[Gamma-rays, annihilation and radio in XRPs]
{
Gamma-ray lines, electron–positron annihilation, and possible radio emission in X-ray pulsars
}
\author[A.~Mushtukov et al.] 
{
Alexander~A.~Mushtukov$^{1}$\thanks{E-mail: 
alexander.mushtukov@physics.ox.ac.uk (AAM)},
Emir~Tataroglu$^{2}$
\thanks{E-mail: 
emirtataroglu2008@gmail.com (ET)},
Alex~J.~Cooper$^{1}$,
Sergey~S.~Tsygankov$^{3}$
\\ 
$^1$ Astrophysics, Department of Physics, University of Oxford, Denys Wilkinson Building, Keble Road, Oxford OX1 3RH, UK\\
$^2$ Department of Physics, University of Virginia, 382 McCormick Rd, Charlottesville, VA, USA \\ 
$^3$ Department of Physics and Astronomy,  FI-20014 University of Turku, Finland \\ 
} 

\pubyear{2025}

\begin{document}
\label{firstpage}
\pagerange{\pageref{firstpage}--\pageref{lastpage}}
\maketitle


\begin{abstract} 
Accretion onto neutron stars (NSs) in X-ray pulsars (XRPs) results in intense X-ray emission, and under specific conditions, high-energy nuclear interactions that produce gamma-ray photons at discrete energies. 
These interactions are enabled by the high free-fall velocities of accreting nuclei near the NS surface and give rise to characteristic gamma-ray lines, notably at 2.2 MeV, 5.5 MeV, and 67.5 MeV. 
We investigate the production mechanisms of these lines and estimate the resulting gamma-ray luminosities, accounting for the suppression effects of radiative deceleration in bright XRPs and the creation of electron-positron pairs in strong magnetic fields. 
The resulting annihilation of these pairs leads to a secondary emission line at $\sim 511$\,keV. 
We also discuss the possibility that non-stationary pair creation in the polar cap region could drive coherent radio emission, though its detectability in accreting systems remains uncertain.
Using a numerical framework incorporating general relativistic light bending and magnetic absorption, we compute the escape fraction of photons and distinguish between actual and apparent gamma-ray luminosities. 
Our results identify the parameter space - defined by magnetic field strength, accretion luminosity, and NS compactness - where these gamma-ray signatures may be observable by upcoming MeV gamma-ray missions. 
In particular, we highlight the diagnostic potential of detecting gravitationally redshifted gamma-ray lines and annihilation features for probing the mass-radius relation and magnetospheric structure of NSs.
\end{abstract}


\section{Introduction}
\label{sec:Intro}

Accretion onto neutron stars (NSs) in close binary systems results in significant acceleration of accreting material in the gravitational field of a star, effective energy release at the NS surface and appearance of a bright source in X-ray sky.
The free fall velocity at the NS surface is expected to be $\sim 0.5c$.
Under this condition, the accretion efficiency can reach $\sim 0.2$.
The accretion luminosity can be expressed as:
\beq 
L \simeq 1.9\times 10^{37}\,\left(\frac{\dot{M}}{10^{17}\,{\rm g\,s^{-1}}}\right)\frac{M}{1.4M_\odot}\frac{10^6\,{\rm cm}}{R}\,\,\ergs,
\eeq 
where $\dot{M}$ is the mass accretion rate,
$M$ and $R$ are the mass and radius of the NS respectively.

Accretion onto strongly magnetised NSs results in the phenomenon of X-ray pulsars (XRPs).
The presence of a strong magnetic field, which is typically $\sim 10^{12}\,{\rm G}$ at the NS surface, affects the accretion process by shaping its geometry and influencing the elementary interaction processes between particles (see, e.g., \citealt{2022arXiv220414185M,2006RPPh...69.2631H}). 
In particular, the accretion flow is directed by NS magnetic field into a small regions of area $\sim 10^{9}\,{\rm cm^2}$ located close to the poles of a star. 

High free-fall velocity and thus the energetic collisions of infall nuclei and particles in the NS atmosphere result in nuclear reactions and production of gamma-ray photons in MeV energy band.
A few channels of gamma-ray photon production were proposed in the literature:
\begin{enumerate}[leftmargin=12pt]
\item
Incoming $^4$He nuclei decelerate due to multiple Coulomb collisions with electrons in the NS atmosphere \citep{1995ApJ...438L..99N}.
Collisions with atmospheric protons result in the partial destruction of these nuclei, leading to the release of neutrons and $^3$He:
\beq 
^4{\rm He} + p \longrightarrow {^3{\rm He}} + p + n.
\eeq
The neutrons produced in this process may either be captured by protons, emitting $2.2$~MeV photons
\beq
p + n \longrightarrow {^2{\rm H}} + \gamma_{2.2}
\eeq
or undergo charge exchange with $^3$He. 
The emission of $2.2$~MeV photons primarily occurs from neutrons that exceed the amount of $^3$He produced, as most $^3$He will capture a neutron, leading to the formation of protons and $^3$H \citep{Bildsten93}.
\item
Proton capture by deuterium ($^2$H) results in the production of $\gamma$-ray photons with energies of $\sim 5.5$~MeV:
\beq\label{eq:2H_p}
{^2{\rm H}} + p \longrightarrow {^3{\rm He}} + \gamma_{5.5}.
\eeq
\item
Proton capture by tritium ($^3$H) produces $\gamma$-rays with energies around $19.8$~MeV \citep[see][and references therein]{Bildsten93,2024A&A...690A.309D}:
\beq\label{eq:tritium}
{^3{\rm H}} + p \longrightarrow {^4{\rm He}} + \gamma_{19.8}.
\eeq
\item 
A small fraction of proton-proton collisions in the atmosphere result in production of neutral pions:
\beq\label{eq:pp-pppi0}
p + p \longrightarrow p + p + \pi^0.
\eeq
The latter tend to decay with production of two gamma-ray photons of energy $\sim 67.5\,{\rm MeV}$ \citep{1970Ap......6...56S}: 
\beq
\pi^0 \longrightarrow 2\,\gamma_{67.5}.
\eeq
\item
A fraction of gamma-ray photons produced in nuclear reactions create electron-positron pairs \citep{2025JHEAp..4800420T} due to one-photon and two-proton pair production
\beq
\gamma \longrightarrow e^- + e^+,\quad\quad
\gamma + \gamma \longrightarrow e^- + e^+.
\eeq
Further annihilation of pairs results in emission of photons at energy $\sim 511\,{\rm keV}$.\\
\end{enumerate}
All of these reactions require high-energy particle collisions \citep{Bildsten93}. However, in the case of bright XRPs, the accretion flow above the NS surface can be decelerated by radiative forces, which reduces the efficiency of these collisions. At the critical luminosity, $L_{\rm crit} \sim 10^{37}\,\ergs$ \citep{1976MNRAS.175..395B,2015MNRAS.447.1847M}, the flow is completely halted by a radiation-dominated shock. Beyond this point, further increases in the mass accretion rate and luminosity lead to the formation of magnetically confined, radiation-supported accretion columns \citep{1981A&A....93..255W,2015MNRAS.454.2539M,2023MNRAS.524.2431S}.
In such conditions, the energy of collisions between heavy particles is significantly reduced, which leads to a corresponding decrease in the luminosity of gamma-ray lines, except for those produced in thermonuclear reactions at the NS surface \citep{2024A&A...690A.309D}.

The total luminosity of XRP in gamma ray lines can be estimated from the efficiency of gamma photon production at the NS surface (see Section\,\ref{sec:PhotonProd}) and photon absorption in the magnetosphere of a star (see Section\,\ref{sec:PhotonAbs}).
However, gamma ray emission tents to be strongly anisotropic: the most of photons escape NS magnetosphere within two cones directed along NS magnetic field axis.
NS rotation should result in pulsations and possible difference between intrinsic and apparent luminosity of an object (see, e.g., \citealt{2024MNRAS.527.5374M}).
To get the relation between actual and apparent luminosity we have to know geometry of NS rotation in the observer's reference frame.
This information can be obtained from observations of pulse phase dependence of X-ray polarisation \citep{2022NatAs...6.1433D,2024Galax..12...46P}.
This method, however, requires sufficiently high X-ray energy flux and was applied by IXPE to a limited number of XRPs: Her~X-1 \citep{2022NatAs...6.1433D,2024NatAs...8.1047H}, Cen~X-3 \citep{2022ApJ...941L..14T}, Vela~X-1 \citep{2023ApJ...947L..20F}, 
X~Persei \citep{2023MNRAS.524.2004M}, 
RX J0440.9+4431 \citep{2023A&A...677A..57D}, 
GRO~J1008-57 \citep{2023A&A...675A..48T}, 
EXO~2030+375 \citep{2023A&A...675A..29M}, 
GX~301-2 \citep{2023A&A...678A.119S}, 
Swift~J0243.6+6124 \citep{2024ApJ...971L..21M,2024arXiv240508107P},
SMC~X-1 \citep{2024A&A...691A.216F}. 

In this paper, we investigate the conditions under which gamma-ray photons produced at the surface of accreting NSs can escape the magnetosphere without being absorbed via one-photon pair production. We develop a numerical framework to compute photon escape fractions, taking into account general relativistic light bending and magnetic absorption effects. 
Based on these calculations, we estimate both the intrinsic and apparent gamma-ray luminosities in specific nuclear lines and evaluate the expected strength of the associated annihilation line at 511\,keV. 
The results are presented for a range of NS magnetic fields and accretion luminosities, highlighting the parameter space where gamma-ray signatures may be observable. 
The structure of the paper is as follows: Section~\ref{sec:Model} introduces the physical setup and relevant processes; Section~\ref{sec:Num} describes the numerical scheme; Section~\ref{sec:NumRes} presents our results; and Section~\ref{sec:Summary} offers a discussion of the implications and observational prospects.

\section{Model set up}
\label{sec:Model}

We consider a case of accretion onto the surface of spherically symmetric magnetised NS assuming that the geometry of space is described by the Schwarzschild metric.

We assume magnetic field dominated by the dipole component.
In this case, the local field strength is given by
\beq
B = \frac{1}{2}\frac{\mu}{r^3}\sqrt{1+3\cos^2\theta_B},
\eeq 
where $\mu=B_0 R^3$ is the magnetic dipole moment, 
$B_0$ is the field strength at the pole of a NS, 
$R$ is NS radius, 
$r$ is the radial distance and $\theta_B$ is the co-latitude in the reference frame of the magnetic dipole.
The unit vector describing local direction of the field lines is
\beq
\mathbfit{n}_b = 
\left(\begin{array}{c} 
-\cos(\chi-\lambda_B)\cos\varphi_B \\ 
-\cos(\chi-\lambda_B)\sin\varphi_B \\
\sin(\chi-\lambda_B)
\end {array}\right),
\eeq 
where $\chi={\rm atan}[0.5\,{\rm tan}\theta_B]$,
$\lambda_B=\pi/2 - \theta_B$ is the latitude, and $\varphi_B$ is the azimuthal angle in the reference frame of NS magnetic dipole.

Accretion from the stellar wind or accretion disc is directed by NS magnetic field towards small regions of area $\sim 10^{9}\,{\rm cm^2}$ located close magnetic poles of a NS.

\subsection{Production of gamma photons}
\label{sec:PhotonProd}

\subsubsection{Proton-neutron recombination: $2.2\,{\rm MeV}$}

The luminosity in $2.2\,{\rm MeV}$ gamma-ray line at the NS surface is dependent on the mass accretion rate and the fraction of $^4$He in the flow, and can be estimated as 
\beq \label{eq:L_2.2}
\frac{L_{2.2}}{L_{\rm X}} \simeq E_{2.2} Q_{2.2}\left(\frac{Y}{4}\right)
\frac{ R_{\rm NS}}{GM_{\rm NS}m_{\rm p}},
\eeq
where $Y$ is the mass fraction of $^4$He atoms in the accretion flow, 
$Q$ is the number of unscattered $2.2\,{\rm MeV}$ photons escaping NS atmosphere per one atom of accreted helium, and $L_{\rm X}=GM_{\rm NS}\dot{M}/R_{\rm NS}$ is the total X-ray luminosity,
$m_{\rm p}$ is proton mass.

Dimensionless factor $Q_{2.2}$ is expected to be dependent on velocity of accretion flow right above NS surface (see Table~4 in \citealt{Bildsten93}) and can be approximated as
\beq\label{eq:Q_2.2}
Q_{2.2}\approx 0.04\,\exp\left[ -\frac{300}{(E_{\rm p}/1\,{\rm MeV})^{1.1}} \right],
\eeq 
where 
\beq\label{eq:E_p_kin}
E_{\rm p} = (1-\gamma_p)m_{\rm p}c^2
\eeq 
is kinetic energy of a proton and $\gamma_{\rm p}=[1-(v/c)^2]^{-1/2}$ is the Lorentz factor.

\subsubsection{Proton capture by deuterium: $5.5\,{\rm MeV}$}

Proton capture by deuterium caused by plasma deceleration in the atmosphere of a NS is less efficient than the process of proton-neutron recombination \citep{Bildsten93}.
The $Q$-factor can be roughly approximated as (see Table~4 in \citealt{Bildsten93}):
\beq
Q_{5.5} \thickapprox 10^{2\ln({E}/{31}) - 9}.
\eeq 

Due to the small efficiency of the process (\ref{eq:2H_p}), the luminosity in 5.5\,MeV line caused by accretion flow deceleration in the atmosphere is expected to be well below the luminosity in 2.2\,MeV line. 
However, high mass accretion rates onto a small regions at the surface of a NS in XRPs resutls in stable nuclear burning of accreting material \citep{1997ApJ...477..897B}, which includes transformation of hydrogen into helium in proton-proton cycle. 
The proton-proton cycle involves the capture of a proton by deuterium, leading to the production of a 5.5 MeV photon.
The luminosity of gamma-ray line in this case can be as high as 
\beq \label{eq:L_5.5}
\frac{L_{5.5}^{\rm (th)}}{L_{\rm X}} \simeq E_{5.5} \left(\frac{X}{2}\right)
\frac{ R_{\rm NS}}{GM_{\rm NS}m_{\rm p}}
\approx 1.4\times 10^{-2} {X},
\eeq 
where $X$ is the mass fraction of hydrogen in the accretion flow. 
The number of gamma photons generated through this process can surpass those produced by plasma deceleration. 
The luminosity estimated in (\ref{eq:L_5.5}) is not affected by accretion flow deceleration in bright XRPs.

{
However, stable nuclear burning of accreted material at high mass accretion rates in XRPs is expected to occur at very large optical depths, where any produced gamma-ray photons are unlikely to escape. 
In this sense, the luminosity given by Eq.~(\ref{eq:L_5.5}) should be regarded as an upper limit, and the actual contribution from stable thermonuclear burning is expected to be much smaller. 
Some burning may still take place in the outer atmospheric layers while the plasma is decelerated, but a realistic estimate of the resulting line luminosity would require a self-consistent treatment of the atmosphere structure, plasma deceleration, and nuclear burning, which is beyond the scope of this work. 
In addition, due to the potential abundance of carbon in the atmosphere of a NS in XRPs, hydrogen may primarily be converted into helium via the CNO cycle. 
Therefore, accurate estimates of photon production at $5.5\,{\rm MeV}$ require solving the kinetic equations that describe nuclear burning in the stellar surface layers.
}

\subsubsection{Proton capture by tritium: $19.8\,{\rm MeV}$}

The proton capture reaction on tritium (\ref{eq:tritium}) produces a gamma-ray photon with energy $E_\gamma = 19.8$\,MeV in the center-of-mass frame. 
While energetically favorable, this reaction is highly suppressed in typical NS atmospheres due to the extremely low abundance of tritium.

The efficiency of this channel is encoded in the $Q$ factor (defined as the number of escaping photons per accreted baryon), which can be approximated as
\beq
Q_{19.8} \thickapprox 10^{1.31\ln(0.025E) - 7},
\eeq
where $E$ is the kinetic energy of the accreting protons in MeV (see Table 4 in \citealt{Bildsten93}). 
The exponential suppression at sub-relativistic energies makes this process highly inefficient for typical accretion velocities.

Moreover, tritium is not expected to be present in significant quantities in freshly accreted material, unless it is produced in situ via nuclear spallation or rare fusion pathways. Even in this case, its rapid destruction in further reactions and low equilibrium abundance severely limits its potential contribution to the gamma-ray line emission.

Given these constraints, the luminosity in the $19.8\,\mathrm{MeV}$ line is expected to be negligible in typical accreting NS environments. A detailed treatment of the relevant reaction networks and equilibrium abundances is beyond the scope of this work and will require a full nuclear kinetics model including spallation and diffusion.

\subsubsection{Neutral pion decay: $67.5\,{\rm MeV}$}
\label{sec:pion_decay}

Protons falling onto a NS surface in accreting XRPs may reach velocities up to $v \sim 0.5c$, corresponding to kinetic energies of approximately $E_k \sim 145$\,MeV. 
While the dominant energy loss mechanism for such protons is Coulomb scattering with ambient electrons, a small fraction of them may undergo inelastic proton-proton ($pp$) collisions, producing neutral pions.

\begin{figure}
\centering 
\includegraphics[width=8.3cm]{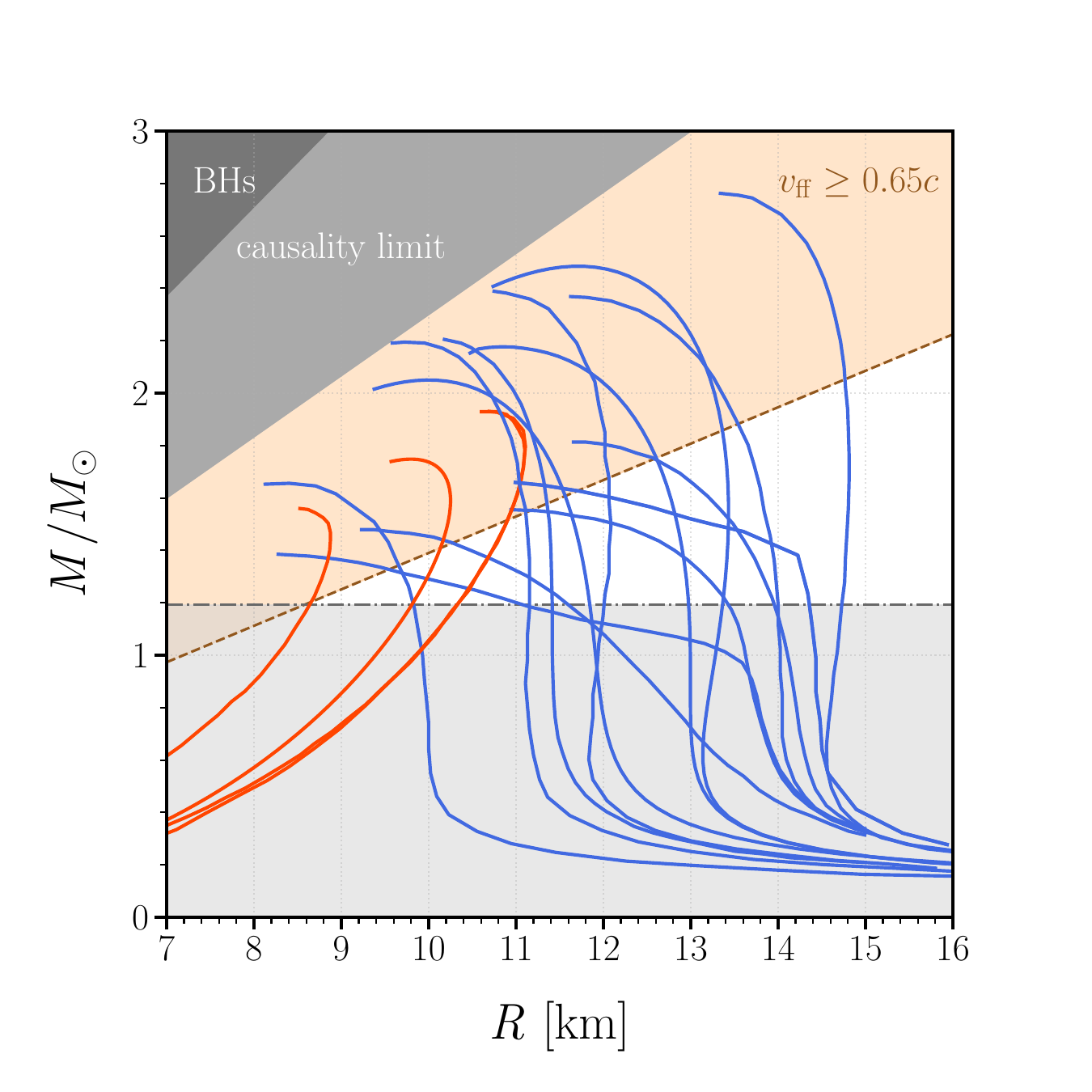}
\caption{
Mass-radius relation for NS calculated under the assumption of different EoS. 
Blue lines correspond to NSs with various EoS of dense matter, while red lines show mass-radius relation for strange stars (see e.g. \citealt{2001ApJ...550..426L}).
The region colored in orange corresponds the combinations of NS mass and radius, when the free-fall velocity $v_{\rm ff}\geq 0.65c$ and the kinetic energy of accreting protons is high enough to cause production of neutral pions.
Dashed-dotted horizontal line shows the minimal mass of a NS that was possible to get in simulations of supernova explosion: $M\sim 1.192\,M_\odot$ \citep{2025PhRvL.134g1403M}.
}
\label{pic:sc_EoS_}
\end{figure}

It is important to note that the reaction (\ref{eq:pp-pppi0}) is a threshold process: it requires the kinetic energy in the laboratory frame to exceed $\sim292$\,MeV. 
Thus, for $pp$ pion production to occur, the infalling protons must acquire sufficient energy during free-fall. 
This condition can be satisfied only if the gravitational potential of the NS is deep enough, i.e., for sufficiently high stellar mass. 
For a canonical NS radius $R = 10$\,km, the required mass is $M \gtrsim 2.1\,M_\odot$. 
As illustrated in Fig.~\ref{pic:sc_EoS_}, the free-fall velocity, and hence the efficiency of pion production, is strongly dependent on the stellar compactness. More compact stars allow infalling protons to reach higher kinetic energies, making pion production more efficient, whereas for lower compactness the threshold is not reached.

The probability of a single proton undergoing a $\pi^0$-producing collision while being stopped in the atmosphere can be estimated using the atmospheric column density $\Sigma$ required to decelerate accretion flow from free-fall velocity to $v\sim 0.64c$ when the kinetic energy of protons drops down to the threshold energy, and the total cross-section for pion production, $\sigma_{\pi^0}$. The number of target nucleons per unit mass is $1/m_p$, where $m_p$ is the proton mass. Therefore, the number of $\pi^0$ mesons produced per unit time is:
\beq 
\dot{N}_{\pi^0} = \sigma_{\pi^0} \frac{\dot{M}}{m_p}  \frac{\Sigma}{m_p},
\eeq
where the typical values for the cross-section slightly above threshold are $\sigma_{\pi^0} \sim 10^{-30}$\,cm$^2$ \citep{2000PhRvD..62i4030B,2007PhRvD..75c4001N}.

Protons lose their kinetic energy $E_{\rm kin}$ mostly in proton-electron collisions according to 
\beq 
\left|\frac{\d E_{\rm kin}}{\d\Sigma}\right| \simeq
\frac{4\pi e^4}{m_e c^2}
\frac{Z^2}{\beta^2}\frac{Z_{\rm pl}}{A}\ln\Lambda,
\eeq 
where $\beta=v/c$ is dimensionless velocity, $Z=1$ for protons, $Z_{\rm pl}$ is the atomic charge number of ions in the atmospheric plasma, $A$ is the atomic mass number of the ions, $\ln\Lambda$ is the Coulomb logarithm.
In non-magnertic plasma, $\ln\Lambda\in[20;40]$ \citep{1969SvA....13..175Z,2005ppa..book.....K}, while in strongly magnetised plasma the Coulomb logarithm is expected to be smaller and $\ln\Lambda<10$ \citep{1980SvA....24..303Y,1999A&A...351..787P}.
The column density required to plasma brake down to velocity $\sim 0.64c$ can be estimated as
\beq 
\Sigma = \frac{\Delta E_{\rm kin}}{\left|\d E_{\rm kin}/\d\Sigma\right|} \simeq 3\times 10^3\,\frac{\beta^2_{\rm ff}\left[\gamma(v_{\rm ff}) - 1.311\right]}{\ln\Lambda}\,\,{\rm g\,cm^{-2}}.
\eeq 
In the case of $v_{\rm ff}=0.7c$ and $\ln\Lambda = 5$, the column density is $\Sigma\simeq 25\,{\rm g\,cm^{-2}}$.

Each $\pi^0$ decays almost instantaneously into two photons with energies $\sim 67.5$\,MeV in the pion rest frame. 
This mechanism may produce a $\gamma$-ray signature in the $\sim$70\,MeV range. 
The luminosity in this $\gamma$-ray line is expected to scale with the X-ray luminosity as
\beq 
\frac{L_{67.5}}{L_X} \approx 135\,\mathrm{MeV}\,\frac{\sigma_{\pi^0} \Sigma}{m_p^2} \frac{R}{GM}
\approx 3.85\times 10^{-7}\,\Sigma\,\frac{R_6}{M_{1.4}}.
\eeq
Thus, while the efficiency of this pion-decay channel is generally low, it may become relevant in systems with high-mass NSs, potentially contributing characteristic line feature in the gamma-ray spectrum. 
Additionally, the luminosity in this line should be sensitive to the deceleration of the accretion flow due to radiative forces, which can significantly reduce the kinetic energy of infalling material in bright XRPs and suppress $\pi^0$ production.

\subsubsection{Influence of radiative force}

\begin{figure}
\centering 
\includegraphics[width=8.5cm]{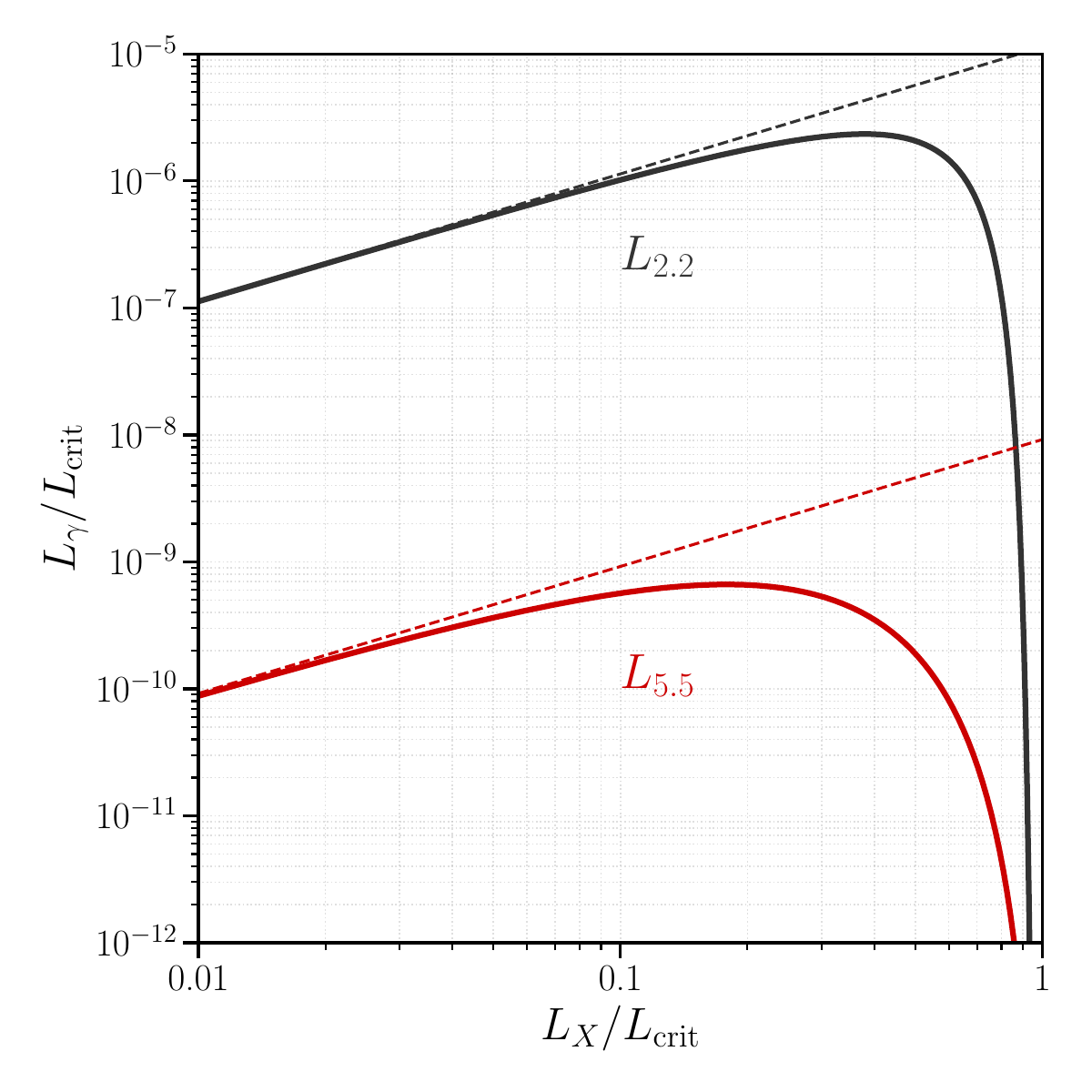}
\caption{
The expected original luminosity in the 2.2 MeV (black) and 5.5 MeV (red ) lines as functions of the total accretion luminosity of X-ray pulsars, $L_X$. Both axes are normalized by the critical accretion luminosity, $L_{\rm crit}$. At low mass accretion rates and luminosities, the gamma-ray lines flux scale proportionally with the total accretion luminosity. However, at high mass accretion rates, radiative forces slow the accretion flow near the NS surface, causing a sharp decrease in the lines luminosity.
}
\label{pic:sc_2.2MeV}
\end{figure}

At mass accretion rates comparable to the critical luminosity \citep{1976MNRAS.175..395B}, radiative force starts to affect dynamics of the flow and decelerate it. 
At the critical luminosity $L_{\rm crit}\sim 10^{37}\,{\rm erg\,s^{-1}}$, which is expected to be dependent on the field strength at the NS surface \citep{2015MNRAS.447.1847M}, the flow is completely decelerated due to its interaction with X-ray photons.
At luminosities, below the critical value, accretion flow velocity can be roughly approximated as 
\beq \label{eq:v}
v \approx v_{\rm ff}\left(1 - \frac{L}{L_{\rm crit}}\right)^{1/2}, 
\eeq 
where 
\beq 
v_{\rm ff} \simeq c\left(\frac{R_{\rm Sh}}{R}\right)^{1/2}
\eeq 
is the free-fall velocity and $R_{\rm Sh}=3\times 10^5\,(M/M_\odot)\,{\rm cm}$ (see Section\,2 in \citealt{2015MNRAS.454.2714M}). 
By combining the dependence of plasma velocity on the luminosity of XRP and the expected relation between the $Q$-factor and proton kinetic energy, one can estimate the total luminosity in the gamma-ray line (see Fig.\,\ref{pic:sc_2.2MeV}). At luminosities well below $L_{\rm crit}$, the gamma-ray line luminosity is proportional to the total accretion luminosity. However, as the total luminosity approaches $L_{\rm crit}$, the gamma-ray line luminosity drops sharply due to the deceleration of the accretion flow by radiative forces rather than Coulomb collisions in the atmosphere (see \citealt{2024A&A...690A.309D}).

\subsection{Photon propagation affected by light bending}

In the case of Schwarzschild metric, the trajectory of each photon lies in one and the same plane. 
Within the plane, photon trajectory can be represented in polar coordinates as a function $r(\varphi)$ that satisfies the differential equation (see Chapter 25 in \citealt{1973grav.book.....M}):
\beq\label{eq:DE_ph_traj}
\frac{\d^2 u}{\d\varphi^2} = 3u^2 - u,
\eeq
where $u \equiv 0.5 R_{\rm Sh}/r$ and $R_{\rm Sh}=2GM_{\rm NS}/c^2$ is the gravitational radius of a NS.
To get a specific trajectory of a photon using (\ref{eq:DE_ph_traj}) one has to specify the coordinate where the photon is emitted and the initial direction of the photon motion. 
While photons move along their trajectories, they pass the regions of different magnetic field strength and change their direction with respect to the local direction of magnetic field.

\subsection{Photon absorption due to pair production}
\label{sec:PhotonAbs}

Gamma-ray photons that are produced at the NS surface and absorbed in the magnetosphere of a NS due to the pair-creation processes give rise to emission in the line at $\sim511\,{\rm keV}$ due to the annihilation of electron-positron pairs \citep[see also][]{1992ApJ...384..143B}.
While two-photon pair production occurs in both magnetic and non-magnetic environments, one-photon pair production is specific to cases with extreme magnetic fields. 
Pair production can significantly reduce the luminosity in gamma-ray photons. 
At the same time, it can enhance the luminosity of XRPs in the annihilation line, as the pairs created in these processes eventually annihilate, producing photon pairs with energy $\sim 511\,{\rm keV}$.

\subsubsection{One-photon pair creation}

One-photon pair creation is forbidden in the field absence case, where it violates the laws of conservation of momentum and energy of particles, but it is possible in strong magnetic fields, where it is enough to fulfill the law of conservation of energy and momentum along the field direction.
A photon propagating at an angle $\theta$ to the local magnetic field can undergo pair production if its energy
\beq 
E \geq 2mc^2\sin\theta.
\eeq 

The attenuation coefficient due to one-photon pair production is dependent on photon energy, magnetic field strength, momentum of photon with respect to the $B$-field direction and photon polarisation state.
The accurate expressions for the attenuation coefficient were derived by \citealt{1983ApJ...273..761D} (see Appendix\,\ref{app:alpha_pp} and red line in Fig.\,\ref{pic:alpha_pp}).
The useful approximations of expression were proposed by \citealt{1966RvMP...38..626E,1988MNRAS.235...51B,2014ApJ...790...61S,2019MNRAS.486.3327H}. 
In this paper, we use attenuation coefficient that is averaged over the polarization state of a photon.

In the case of 
\beq
\frac{1}{b}
\left(\frac{E\sin\theta}{mc^2}\right)^2\gg 1
\eeq
where 
$b\equiv B/B_{\rm cr}$ is magnetic field strength in units of critical field strength $B_{\rm cr}\simeq 4.413\times 10^{13}\,{\rm G}$,
the attenuation coefficient due to one-photon pair production can be approximated as \citep{2007PhRvD..75g3009B}
\beq\label{eq:alpha_PP_approx}
\alpha_{1\gamma}(E,b,\theta) = \frac{\alpha}{\lambdabar}b\sin\theta\,\xi(\omega_\perp,b),
\eeq
where $\alpha = 1/137$ is the fine structure constant, 
$\lambdabar$ is the electron Compton wavelength,
$\omega_\perp = E\sin\theta/mc^2$,
\beq 
\xi(\omega_\perp,b) = \frac{3\omega_\perp^2 - 4}{2\omega_\perp^2\sqrt{(\omega_\perp^2-4)\mathcal{L}(\omega_\perp)\phi(\omega_\perp)}},
\eeq 
and 
\beq 
\mathcal{L}(\omega_\perp) &=& \ln\left(\frac{\omega_\perp+2}{\omega_\perp-2}\right),\\ 
\phi(\omega_\perp)&=&4\omega_\perp-(\omega_\perp^2-4)\mathcal{L}(\omega_\perp).
\eeq  
Approximation (\ref{eq:alpha_PP_approx}) does not reproduce multiple resonances that appear in the absorption coefficient, but can be used as rough approximation for magnetic fields $B\lesssim 10^{12}§\,{\rm G}$ (see Fig.\,\ref{pic:alpha_pp}).
In the case of stronger magnetic fields, approximation (\ref{eq:alpha_PP_approx}) underestimates the actual absorption coefficient. 

\begin{figure}
\centering 
\includegraphics[width=8.7cm]{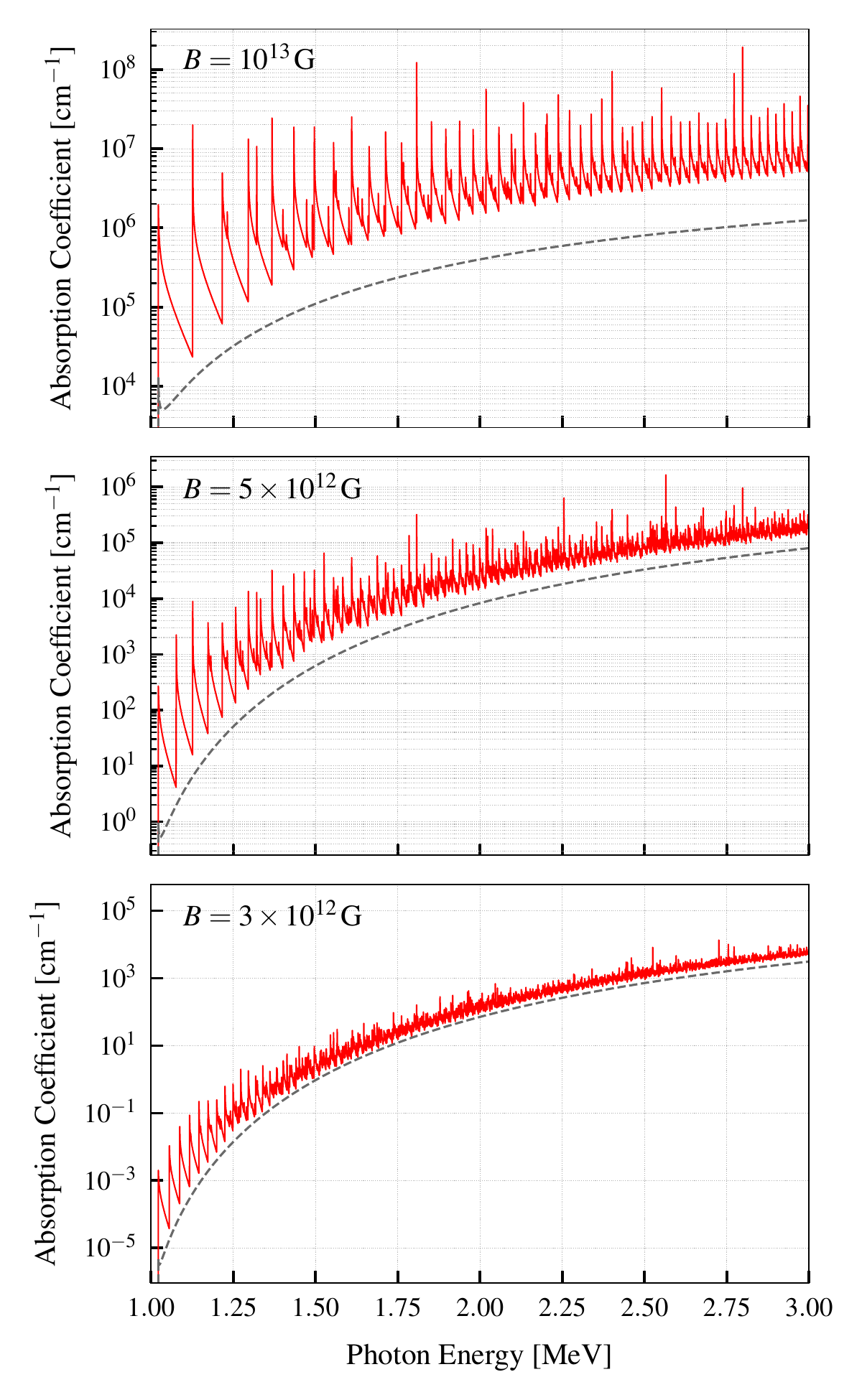}
\caption{
The absorption coefficient (averaged over the polarisation states) due to one-photon pair production in magnetic field as a function of photon's energy.
The absorption coefficients are calculated according to \citealt{1983ApJ...273..761D} for photons propagating across magnetic field lines (i.e., $\theta=\pi/2$, solid red lines).
Approximations (\ref{eq:alpha_PP_approx}) are shown by dashed grey lines. 
The lower, middle and upper panels show the results for the case of magnetic field strength $B=3\times 10^{12}\,{\rm G}$, $5\times 10^{12}\,{\rm G}$, and $10^{13}\,{\rm G}$ respectively.
}
\label{pic:alpha_pp}
\end{figure}

The optical thickness along specific photon trajectory due to one-photon pair creation can be calculated as
\beq\label{eq:tau}
\tau (s_1,s_2) = \int\limits_{s_1}^{s_2}\d s\,\alpha_{1\gamma}(E(s),b(s),\theta(s)),
\eeq 
where we integrate along the trajectory,
\beq 
\d s = \sqrt{ \left(1 - \frac{R_{\rm Sh}}{r(\varphi)} \right)^{-1} \left( \frac{dr}{d\varphi} \right)^2 + r(\varphi)^2 }\, d\varphi,
\eeq 
and
\beq 
E = E_{NS} \sqrt{ \frac{r}{R_{NS}}\frac{R_{NS}-R_{\rm Sh}}{r-R_{\rm Sh}} }
\eeq 
is the local photon energy affected by the gravitational redshift, and 
$\theta = {\rm acos}(\mathbfit{n}_b \cdot \mathbfit{n}_{\rm ph}) $ 
is the angle between local directions of the field given by $\mathbfit{n}_b$ and photons momentum given by $\mathbfit{n}_{\rm ph}$.

The fraction of photons that can path through the magnetosphere of a NS can be estimated as
\beq \label{eq:fraction}
f = e^{-\tau(0,\infty)}\leq 1.
\eeq

\subsubsection{Possibility of coherent radio emission}

{
If a rotationally-powered electric field is present in the polar cap region, and one-photon pair creation occurs in this region, the electric field will be screened, plausibly in a non-stationary manner. The polar cap encompasses the region above a co-latitude of 
$$\theta_{\rm PC} \sim  \left(\frac{2 \pi R}{c P}\right)^{1/2} \approx 1.4 \times 10^{-2} \, P_1^{-1/2},$$
and has an associated area of: 
$A_{\rm pc} = {4 \pi^2 R^3}{P^{-1}c^{-1}} \sim 3 \times 10^{9} \, P^{-1} \, {\rm cm^{2}}$. 
As pair-producing gamma-rays are concentrated on the magnetic poles, it is reasonable to assume a significant fraction of pair creation will occur within this open-field line region. 
If the pair creation screens the electric field in a non-stationary manner, broadband radio emission may be produced \citep{timokhin2013}. 
For rapidly rotating X-ray pulsars ($P \lesssim 1$~s, i.e., those not beyond the deathline, \citealt{chenruderman93}), newly-created pairs will be accelerated to high-energies and produce curvature photons capable of pair production, initiating pair cascades which further screen the field \citep{ruderman1975}. The expected radio luminosity will be proportional to the pair luminosity with some radio efficiency of order $\eta \sim 10^{-4}$ \citep{cooper23}, which in this case will scale with the incident $\gamma$-ray luminosity above the pair creation threshold such that: $L_{\rm r} = \eta L_{\gamma, \rm MeV}$.
}
We note, however, that in accreting systems the magnetic field topology may be substantially modified (see, e.g., \citealt{2014EPJWC..6401001L}). 
Some fraction of the field lines may connect to the accretion disc or the companion star rather than extend to the light cylinder, and the associated plasma inflow could suppress charge starvation in the polar cap region. 
Because of these uncertainties, the presence of open field lines capable of sustaining coherent radio emission is not guaranteed, which makes the detectability of radio emission in such systems highly uncertain.

\subsection{Annihilation line at $511\,{\rm keV}$ and cyclotron emission}

Gamma-ray photons that are produced at the NS surface and absorbed in the magnetosphere of a NS due to the pair-creation processes give rise to emission in the line at $\sim511\,{\rm keV}$ due to the annihilation of electron-positron pairs. 
Assuming that all pairs produced by gamma-ray photons are annihilated, we can estimate the luminosity in the annihilation line from above:
\beq 
L_{0.511} \simeq \sum\limits_{j} \frac{1.022}{E_{j,\rm MeV}}(1-f_j)L_{j}^{\rm (ini)},
\eeq 
where $L_j^{\rm (ini)}$ is the initial luminosity in specific gamma-ray line, $E_{j,\rm MeV}$ is photon's energy in the line in units if MeV, and $f_j$ is the fraction of gamma-ray photons that are able to penetrate through the magnetosphere.

It is expected that a fraction of electrons and positrons are produced at the excited Landau levels. 
Under the conditions expected in right above NS surface, the collisional de-excitation rate is much smaller than the radiative de-excitation rate \citep{1979A&A....78...53B}.
Thus, particles should experience de-excitation emitting cyclotron photons at local cyclotron energy. 

{
One-photon annihilation becomes significant only in ultrastrong fields, transforming the characteristic 511\,keV two-photon annihilation line into a broadened or even continuum-like feature for \(B' \gtrsim 0.3\,B_{\rm cr}\sim10^{13}\,{\rm G}  \) (see fig.\,11 in \citealt{1980ApJ...238..296D} and \citealt{1986JETP...64.1173K}). 
Detailed calculations show that in such strong fields the angular distribution and polarization of two-photon annihilation are heavily modified, especially for particles in the ground Landau level. 
Three-photon annihilation is negligible under these field strengths, and that two-photon processes remain dominant unless fields exceed \(10^{13}\,{\rm G}\).
}



\section{Numerical model}
\label{sec:Num}


To determine the escape fraction of gamma-ray photons emitted near the surface of a NS, we solve numerically the photon trajectory equation (\ref{eq:DE_ph_traj}), following the numerical scheme described in \citet{2024MNRAS.527.5374M}. 
Each photon is initialized with a given energy and propagation direction from one of the magnetic poles of the NS.

As photons propagate through curved spacetime, their energy is affected by gravitational redshift. 
At each point along the trajectory, we compute the angle between the photon momentum and the local magnetic field direction, assuming a dipolar magnetic field structure.
This allows us to evaluate the local absorption coefficient due to one-photon pair production, which depends on the photon energy, magnetic field strength, and propagation angle.
Calculating the absorption coefficient, we follow accurate description of one-photon pair production (see Appendix~\ref{app:alpha_pp}) unless the number of Landau levels required to get the result exceeds 500 and approximation (\ref{eq:alpha_PP_approx}) works well enough.

We then integrate the attenuation coefficient along each photon trajectory to compute the optical depth (\ref{eq:tau}). 
The corresponding escape probability is determined by the exponential attenuation factor (\ref{eq:fraction}). 
By repeating this procedure for many photon trajectories covering a range of emission directions, we build up the angular distribution of escaping photons.

Assuming a specific beam pattern of gamma-ray photon emission at the NS surface, we integrate over all emission directions to obtain the total fraction of photons that escape the magnetosphere without absorption. This quantity depends sensitively on the photon energy, magnetic field strength, and emission geometry, and is used in subsequent calculations of the emergent gamma-ray luminosity and the contribution to the annihilation line at 511\,keV.



\section{Numerical results}
\label{sec:NumRes}

Gamma-ray photons with energies \( E > 2 m_{\rm e} c^2 \) can escape the XRP magnetosphere only within a cone aligned with the magnetic axis of a NS.  
The opening angle of this cone is determined by the photon energy \( E \) (since pair production is allowed only for photons satisfying \( \sin\theta > \sin\theta_{\rm min} = E / 2 m c^2 \)), the magnetic field strength and structure (which strongly affect the absorption coefficient; see Fig.~\ref{pic:alpha_pp}), and the NS mass and radius, which influence photon trajectories and thus the resulting beam pattern.  
Photons emitted outside the cone are fully or partially absorbed via one-photon electron--positron pair creation.

\begin{figure}
\centering 
\includegraphics[width=8.5cm]{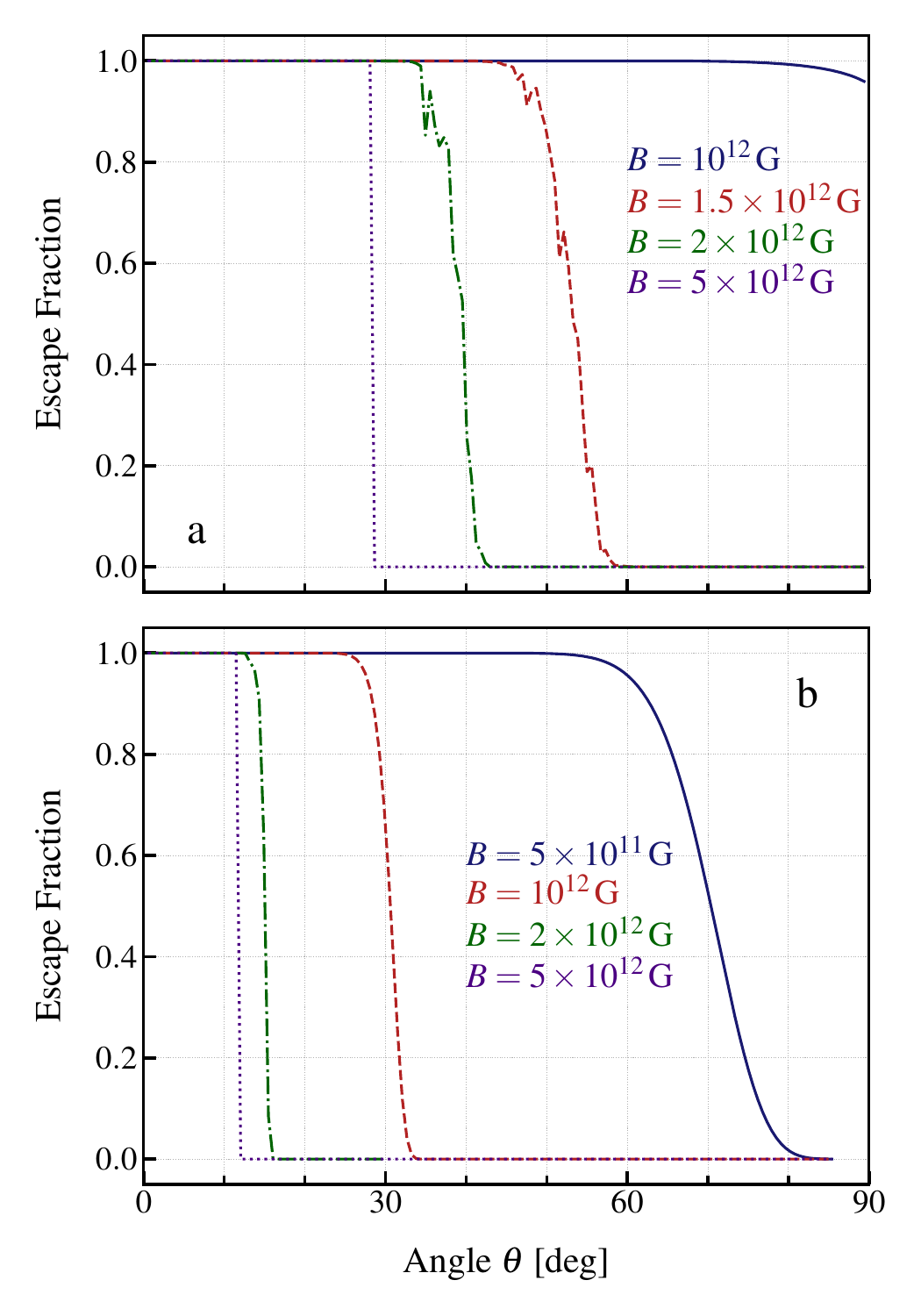}
\caption{
The dependence of the fraction of photons that can escape through the magnetosphere without absorption (due to one-photon pair creation) as a function of their initial emission angle at the NS surface (measured from the magnetic axis).
The upper panel shows results calculated for the case of $E=2.2\,{\rm MeV}$ at the NS surface, while the lower panel shows results for the case of $E=5.5\,{\rm MeV}$.
Different curves show the escape fraction calculated for different magnetic field strength at the stellar surface.
Upper panel:
$10^{12}\,{\rm G}$ (solid line),
$1.5\times 10^{12}\,{\rm G}$ (dashed),
$2\times 10^{12}\,{\rm G}$ (dashed-dotted),
and
$5\times 10^{12}\,{\rm G}$ (dotted).
Lower panel:
$5\times 10^{11}\,{\rm G}$ (solid line),
$10^{12}\,{\rm G}$ (dashed),
$2\times 10^{12}\,{\rm G}$ (dashed-dotted),
and
$5\times 10^{12}\,{\rm G}$ (dotted).
Parameters: $M=1.4\,M_\odot$, $R=10\,{\rm km}$, dipole magnetic field structure.
}
\label{pic:escape_fr}
\end{figure}

In general, stronger magnetic fields and higher photon energies result in smaller escape cone opening angles (see Fig.~\ref{pic:escape_fr}).  
For photons with energies of \( 2.2\,\mathrm{MeV} \) (\( 5.5\,\mathrm{MeV} \)), the NS magnetosphere remains transparent only for surface fields \( B < 10^{12}\,\mathrm{G} \) (\( B < 10^{11}\,\mathrm{G} \)).  
For surface field strengths \( B > 5 \times 10^{12}\,\mathrm{G} \), the opening angle of the escape cone stabilizes at approximately \( 30\degr \) (\( 10\degr \)) for \( 2.2\,\mathrm{MeV} \) (\( 5.5\,\mathrm{MeV} \)) photons (compare the upper and lower panels in Fig.~\ref{pic:escape_fr}).  
As shown in Fig.~\ref{pic:escape_fr}, the escape fraction for the 2.2\,MeV line exhibits non-smooth, spiky features at \( \theta > \theta_{\rm min} \). 
These are not numerical artifacts, but arise from resonances in the pair-production cross section (see Fig.~\ref{pic:escape_fr}). 
When photons propagate through the magnetosphere, they cross regions where the resonance condition is satisfied, which modifies the angular distribution of escaping photons. 
Such resonant effects are much weaker for the 5.5\,MeV case, resulting in smoother escape fraction curves.

\begin{figure}
\centering 
\includegraphics[width=8.5cm]{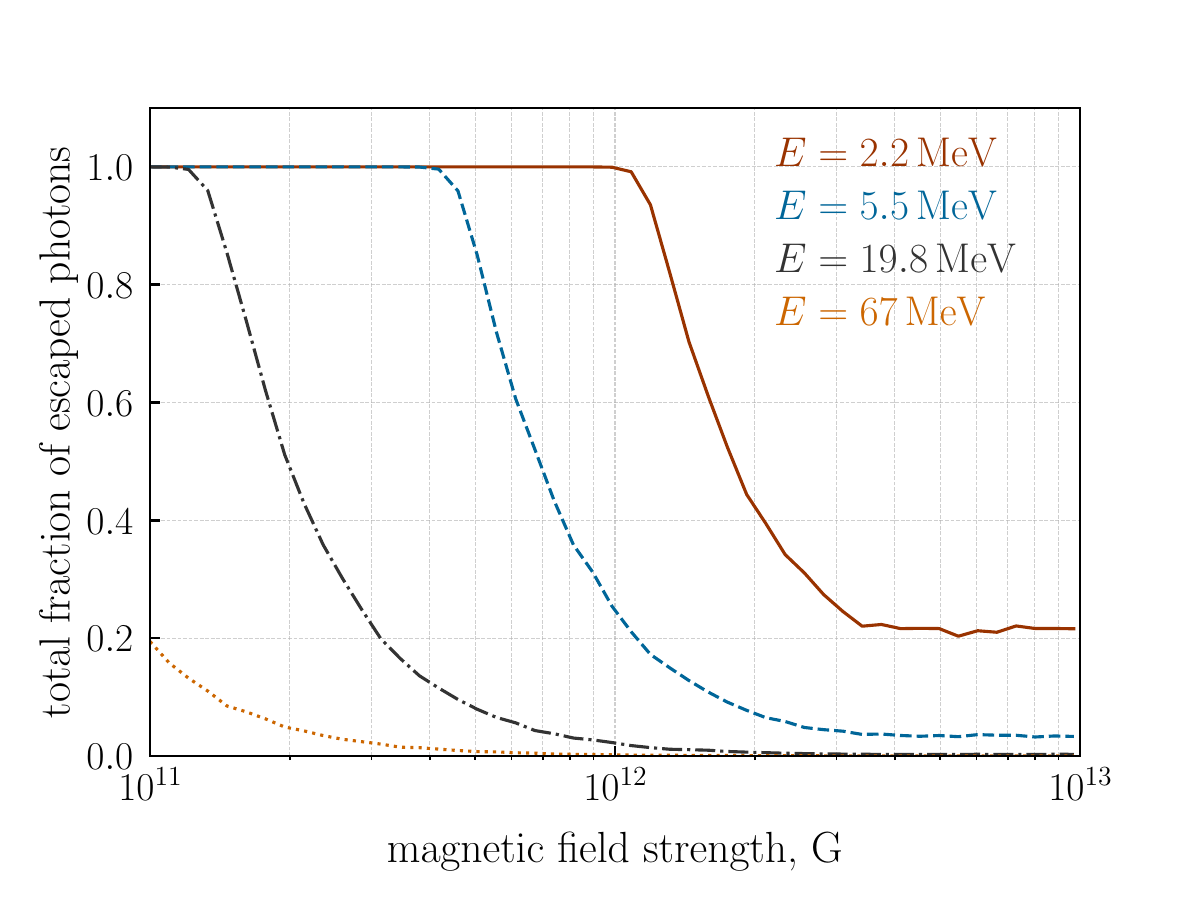}
\caption{
The fraction of photons that penetrate through the magnetosphere of a NS as a function of surface magnetic field strength.
Different lines are given for different energy of gamma-ray photons:
$2.2\,{\rm MeV}$ (solid red),
$5.5\,{\rm MeV}$ (dashed blue),
$19.8\,{\rm MeV}$ (dashed-dotted black),
$67\,{\rm MeV}$ (dotted orange).
}
\label{pic:frac}
\end{figure}

Assuming some specific beam pattern of emitted gamma-ray photons at the magnetic pole of a NS, we get the total fraction of escaped gamma-ray photons (see Fig.\,\ref{pic:frac}).
The fraction of photons that escape the magnetosphere of a NS tend to drop with the increase of surface magnetic field strength. 
In the limit of a very strong magnetic field, the fraction is stabilized because there is always a range of angles, where photons cannot create pairs and escape.
In particular, for $2.2\,{\rm MeV}$ ($5.5\,{\rm MeV}$) photons, the fraction of escaping photons does not drop below $\sim 0.2$ ($\sim 0.03$).

\begin{figure}
\centering 
\includegraphics[width=8.5cm]{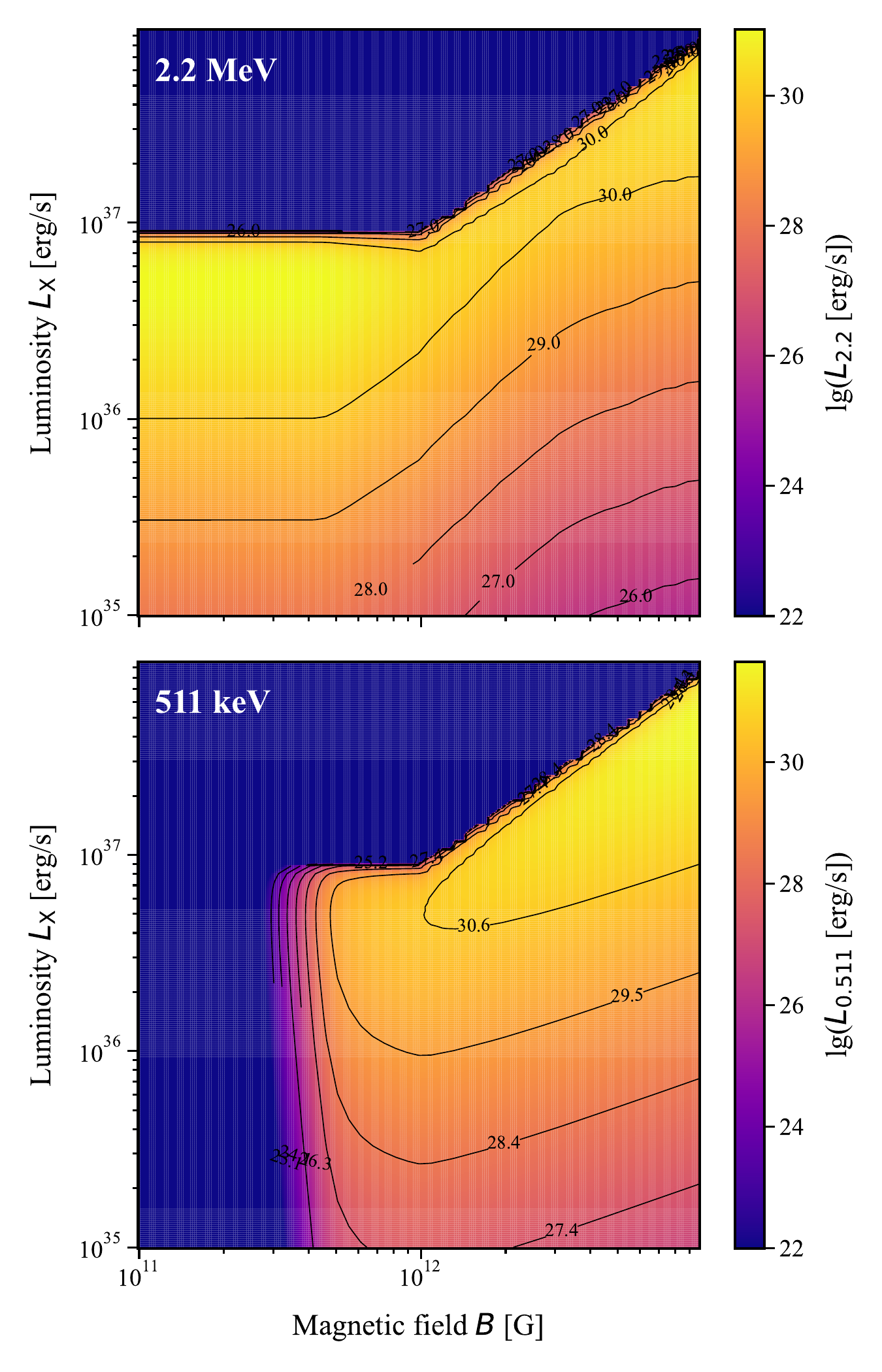}
\caption{
Predicted isotropic luminosities of characteristic gamma-ray lines as a function of NS magnetic field $B$ and X-ray luminosity $L_X$. 
\textbf{Top panel:} Luminosity in the 2.2~MeV deuteron-capture line, accounting for photon attenuation due to one-photon pair production in the magnetosphere. 
\textbf{Bottom panel:} Luminosity in the 511~keV annihilation line, produced by electron-positron pairs generated by escaping 2.2~MeV photons. 
Color indicates $\log_{10}$ of the line luminosity in erg s$^{-1}$. Contours show constant values of $\log_{10}(L_{\mathrm{line}})$.
}
\label{pic:sc_BL_plane}
\end{figure}

\begin{figure}
\centering 
\includegraphics[width=8.5cm]{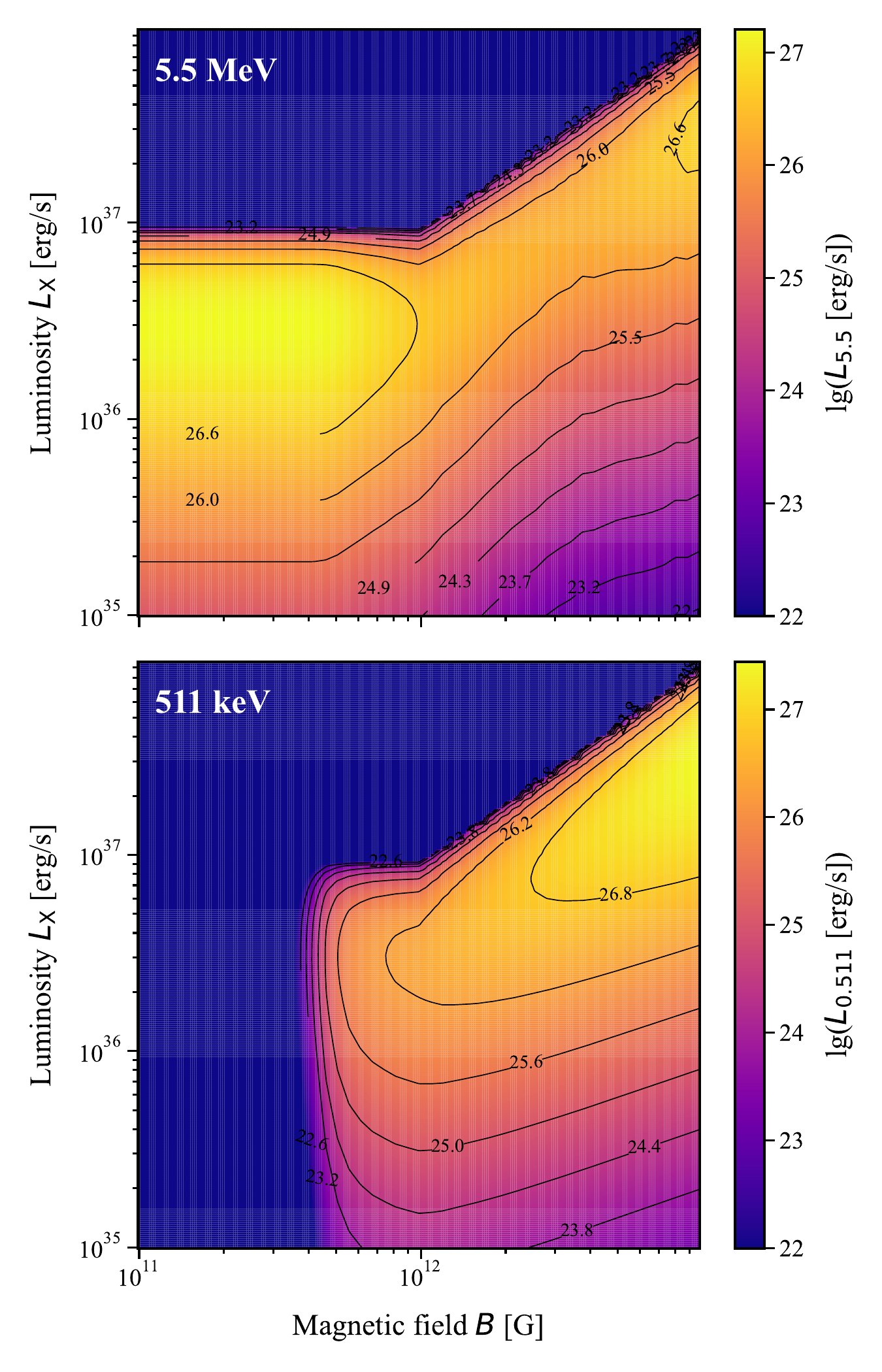}
\caption{
Predicted isotropic luminosities of gamma-ray lines as a function of NS magnetic field $B$ and X-ray luminosity $L_X$. 
\textbf{Top panel:} Luminosity in the 5.5~MeV line, accounting for photon attenuation due to one-photon pair production in the magnetosphere of XRP. 
\textbf{Bottom panel:} Luminosity in the 511~keV annihilation line, produced by electron-positron pairs generated by escaping 5.5~MeV photons. 
Color indicates $\log_{10}$ of the line luminosity in erg s$^{-1}$. Contours show constant values of $\log_{10}(L_{\mathrm{line}})$.
}
\label{pic:sc_BL_plane_5}
\end{figure}

\begin{figure}
\centering 
\includegraphics[width=8.5cm]{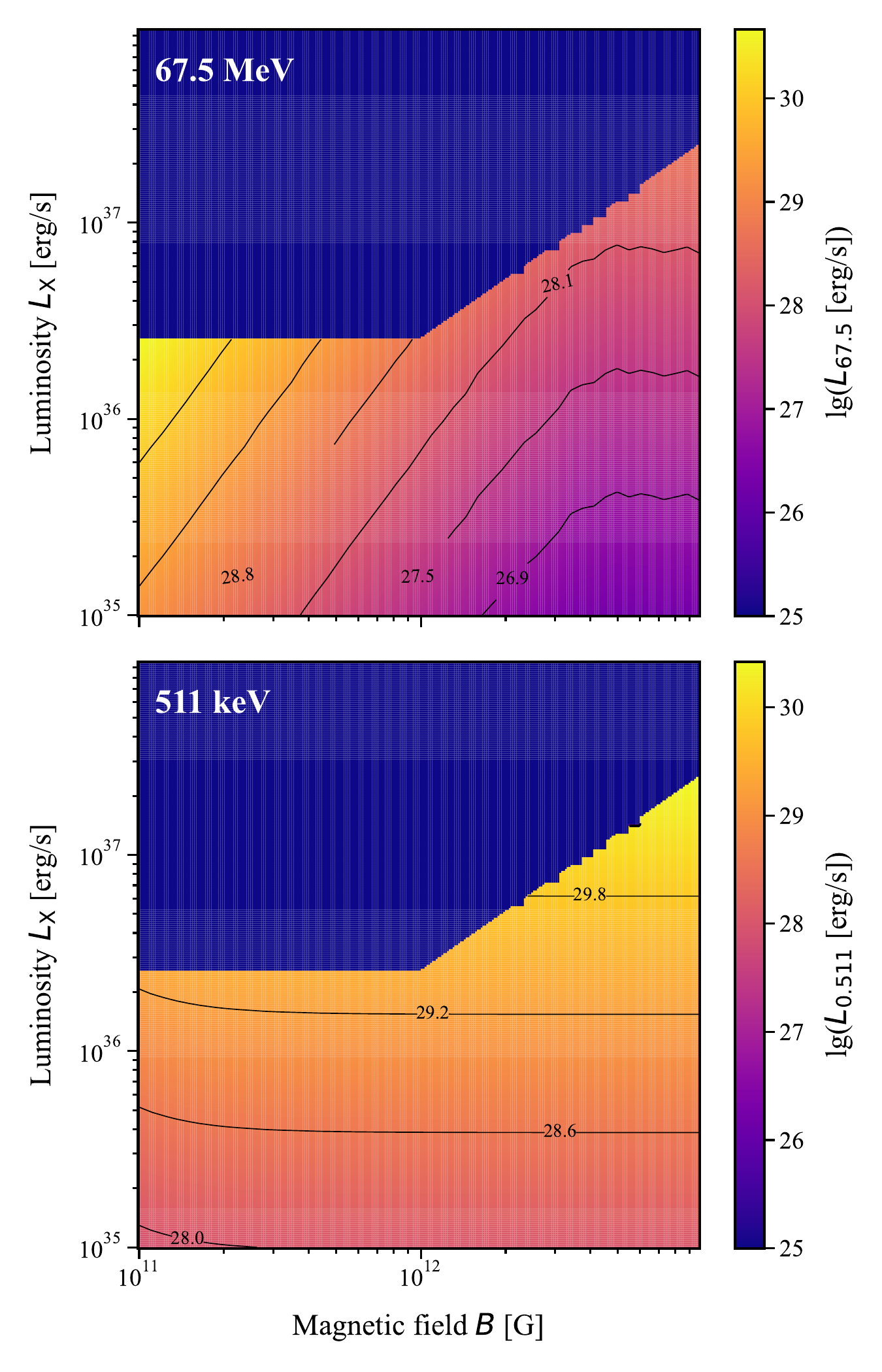}
\caption{
The expected isotropic luminosities of gamma-ray lines as a function of NS magnetic field $B$ and X-ray luminosity $L_X$. 
\textbf{Top panel:} Luminosity in the 67.5~MeV line, accounting for photon attenuation due to one-photon pair production in the magnetosphere. 
\textbf{Bottom panel:} Luminosity in the 511~keV annihilation line, produced by electron-positron pairs generated by escaping 67.5~MeV photons. 
Color indicates $\log_{10}$ of the line luminosity in erg s$^{-1}$. Contours show constant values of $\log_{10}(L_{\mathrm{line}})$.
Parameters: $v_{\rm ff}=0.7c$.
}
\label{pic:sc_BL_plane_67}
\end{figure}

Fig.~\ref{pic:sc_BL_plane} shows isotropic gamma-ray luminosity in the 2.2\,MeV line (upper panel) and the associated luminosity in the 511\,keV annihilation line (lower panel), plotted as a function of NS surface magnetic field strength and X-ray luminosity. 
The gamma-ray luminosity is strongly suppressed at high X-ray luminosities. 
This effect arises due to the deceleration of the accretion flow by radiation pressure at super-critical accretion rates, which reduces the kinetic energy of infalling particles and thus the efficiency of nuclear interactions at the NS surface. 
Simultaneously, the annihilation line luminosity increases in the intermediate regime as more 2.2\,MeV photons are absorbed via one-photon pair production and subsequently converted into electron–positron pairs that annihilate.

A similar behavior is observed in Fig.~\ref{pic:sc_BL_plane_5}, which presents the results for the 5.5\,MeV gamma-ray line. 
Again, the gamma-ray luminosity decreases at high X-ray luminosities due to flow deceleration. 
In this case, however, the transition occurs at slightly different luminosity scales due to the different origin of 5.5\,MeV photons (which can be produced in thermonuclear reactions even when collisional channels are suppressed). 
The corresponding annihilation luminosity again tracks the absorbed gamma-ray power.

Fig.~\ref{pic:sc_BL_plane_67} shows the case of the 67.5\,MeV line, produced through the decay of neutral pions formed in inelastic proton–proton collisions. 
This is a threshold process, requiring extremely energetic collisions, and thus the photon production efficiency is highly sensitive to the free-fall velocity of accreting protons. For this figure, we assume $v_{\rm ff} = 0.7c$, which corresponds to the case of a very compact and massive NS. 
For lower $v_{\rm ff}$, pion production becomes significantly less efficient or even entirely suppressed. 
As a result, both the gamma-ray and annihilation line luminosities are considerably lower or absent altogether for less massive NSs.

In all cases (Fig.~\ref{pic:sc_BL_plane}, \ref{pic:sc_BL_plane_5}, and \ref{pic:sc_BL_plane_67}), we observe that the gamma-ray luminosity is also suppressed at high magnetic field strengths. 
This is due to enhanced one-photon pair creation in strong magnetic fields, which narrows the escape cone of photons and reduces the escaping fraction. This same process leads to increased production of annihilation line photons, since absorbed gamma-ray photons generate pairs that subsequently annihilate. 
Note that in our calculations, we assume that all electron–positron pairs produced via gamma-ray absorption eventually annihilate, which makes our predictions for the 511\,keV luminosity an upper limit.

We emphasize that the relative importance of different gamma-ray and annihilation line contributions is strongly dependent on NS parameters. 
In particular, for massive and compact NSs, the production of 67.5\,MeV photons and their corresponding annihilation signals can be significant and may even compete with or exceed the contribution from lower-energy gamma-ray lines.

\section{Summary and discussion}
\label{sec:Summary}

\subsection{Gamma-ray line emission}

The acceleration onto the NS surface results in emission of photons predominantly in the X-ray energy band. 
Additionally, high-energy collisions between atomic nuclei lead to production of gamma-ray photons at specific energies. 
In particular, photon production is expected in MeV energy band at 
$2.2\,{\rm MeV}$, 
$5.5\,{\rm MeV}$, 
$19.8\,{\rm MeV}$, and 
$67\,{\rm MeV}$.
Because production of gamma-ray photons in these lines requires high-energy collisions, the gamma-ray photon production rate is affected by velocity of accretion flow above NS surface and, thus, by X-ray accretion luminosity, which tends to decelerate accretion flow due to the radiative force \citep{1976MNRAS.175..395B,2015MNRAS.447.1847M}. 
As a result, gamma-ray photon luminosities in all considered lines are suppressed at high $L_{\rm X}$ limit.
The maximal gamma-ray line production is expected at X-ray luminosity $L_{\rm X}\simeq 0.5L_{\rm crit}$.
At $L_{\rm X}\gtrsim L_{\rm crit}$, the production rates drop sharply due to reduced kinetic energy of infalling protons.

Gamma-ray photons produced at $5.5\,{\rm MeV}$ can be an exception from this rule because they are also products of thermonuclear reactions within the proton-proton cycle. 
We expect stable thermonuclear burning at the NS surface in XRPs, and in this case, gamma-ray line production at $5.5\,{\rm MeV}$ is dominated by nuclear burning in the NS atmosphere
or accretion column, even at super-critical accretion rates.

Under the conditions of extremely strong magnetic field, gamma-ray photons produced at the NS surface experience absorption due to the process of one-photon pair creation \citep{1983ApJ...273..761D}. 
Only a fraction of gamma-ray photons can escape the magnetosphere of a NS (see Fig.\,\ref{pic:escape_fr} and \ref{pic:frac}). 
To estimate the escape fraction, we have performed Monte Carlo simulations assuming the Schwarzschild metric and a magnetic field dominated by the dipole component. 
High-energy photons tend to escape in directions close to the magnetic axis of a NS.
The fraction of escaped photons drops with increasing photon energy and magnetic field strength, due to narrowing of the escape cone caused by enhanced pair creation (see Fig.\,\ref{pic:frac}).

The case of $67.5\,{\rm MeV}$ photons is especially sensitive to the compactness of the NS (see Section~\ref{sec:pion_decay}).
This line originates from decay of $\pi^0$ mesons produced in inelastic proton-proton collisions, which occur only above a high-energy threshold.
Therefore, efficient production of $67.5\,{\rm MeV}$ photons requires large free-fall velocity ($v_{\rm ff} \gtrsim 0.65c$), possible only in massive and compact NSs.
At lower $v_{\rm ff}$, this process becomes highly inefficient or does not occur at all.

{
We note that the MeV nuclear lines discussed in this work are expected to be gravitationally redshifted in XRPs. 
If a line with rest-frame energy $E_0$ is observed at $E_{\rm obs}$, the corresponding redshift is 
$1+z = E_0/E_{\rm obs} = (1-2GM/Rc^2)^{-1/2}$. 
As illustrated in Fig.~\ref{pic:sc_EoS_}, this relation directly links the observed line energy to the stellar compactness $M/R$, and therefore to the NS mass–radius relation predicted by different equations of state. 
Thus, even a single secure detection of a redshifted nuclear line would provide a powerful and independent constraint on the compactness of accreting NSs.
}

{
In addition, the detectability of nuclear $\gamma$-ray lines is expected to depend strongly on the rotational geometry of the NS. 
Recent polarimetric measurements with IXPE have shown that X-ray polarization data can constrain the relative orientation of the magnetic and spin axes \citep{2024Galax..12...46P}. 
Combining such geometric constraints with models of photon escape cones enables predictions of the expected pulse profiles in the MeV band. 
These predictions can be directly used to optimize searches with future $\gamma$-ray missions, increasing the likelihood of detection even for weak line fluxes.
}

\subsection{Secondary emission: annihilation line and radio emission}

Gamma-ray photons absorbed by one-photon pair production give rise to photon creation at energy $\sim 511\,{\rm keV}$ via two-photon annihilation of electron-positron pairs which is expected to dominate under magnetic field strengths of
$10^{11}\,{\rm G}\lesssim B \lesssim 10^{13}\,{\rm G}$ \citep{1980ApJ...238..296D,1986JETP...64.1173K}.
Our calculations show that annihilation line luminosity tracks the absorbed fraction of each gamma-ray line and provides an upper limit assuming all pairs annihilate.
In the case of massive NSs with $v_{\rm ff} \sim 0.7c$, the annihilation line produced by absorbed $67.5\,{\rm MeV}$ photons can compete with or exceed that generated by absorption of $2.2\,{\rm MeV}$ and $5.5\,{\rm MeV}$ photons.
In particular, the luminosity in the annihilation line can be comparable to or even exceed that of the escaping primary gamma-ray line, especially at high $B$-field (see Fig.\,\ref{pic:sc_2.2MeV},\,\ref{pic:sc_BL_plane_5} and \ref{pic:sc_BL_plane_67}).

{
In addition to the gamma-ray signatures, we have discussed the possibility that non-stationary pair creation in the polar cap region could drive coherent radio emission. 
While the complex magnetospheric geometry \citep{2014EPJWC..6401001L} in accreting systems may reduce its efficiency, such emission cannot be ruled out. 
We also note that radio emission has been detected from accreting X-ray pulsars such as Her X-1 and Swift J0243.6+6124 \citep{2018MNRAS.473L.141V,2018Natur.562..233V}, although in these cases the radio output is commonly interpreted as being associated with jet activity rather than coherent pulsar emission. 
Nevertheless, a potential detection of pulsed radio signals from XRPs, even if rare, would open a valuable new window into pair cascades and the magnetic field structure in these objects.
}

\section*{Acknowledgements}

AAM thanks UKRI Stephen Hawking fellowship.
The authors thank Dmitry Yakovlev for instructive discussion.
We are also grateful to an anonymous referee for their useful comments and suggestions.

\section*{Data availability}

The calculations presented in this paper were performed using a private code developed and owned by the corresponding author. All the data appearing in the figures are available upon request. 


\appendix

\section{Calculation of Magnetic one-photons Pair Production absorption Coefficients}
\label{app:alpha_pp}

The attenuation coefficients for one-photon magnetic pair production are calculated following the expressions derived by \citet{1983ApJ...273..761D}. The total absorption rate depends on the photon energy \( E \), magnetic field strength \( B \), angle \( \theta \) between local direction of magnetic field and photon momentum, and photon polarization relative to the magnetic field.

Dimensionless photon energy in the electron rest frame is given by
\beq 
\omega' = \frac{E \sin \theta}{2 m c^2}.
\eeq
The absorption coefficient for photons polarized \textit{parallel} to the magnetic field is:
\begin{align}
R_\parallel(\omega', b) = \frac{\alpha b}{2 \lambda_e}
\sum_{j,k} \bigg[ 
& A_{jk}^{(1)} +  A_{jk}^{(2)}
\bigg],
\end{align}
where
\beq 
A_{jk}^{(1)}&=&(E_j E_k + \omega'^2 - p^2)
\left( |M_{j,k}|^2 + |M_{j{-}1,k{-}1}|^2 \right)
\nonumber \\
A_{jk}^{(2)} &=& 2 \sqrt{jk} \, \omega' b\, m^2 
\left[ M_{j,k} M_{j{-}1,k{-}1} + M_{j{-}1,k{-}1} M_{j,k} \right]
\nonumber
\eeq 
while for \textit{perpendicular} polarization, the expression becomes:
\begin{align}
R_\perp(\omega', b) = \frac{\alpha b}{2 \lambda_e}
\sum_{j,k} \Bigg[
A_{jk}^{(3)} - A_{jk}^{(4)}
\Bigg],
\end{align}
where 
\beq 
A_{jk}^{(3)}&=&\left( E_j E_k + \omega'^2 - p^2 \right)
\left( |M_{j,k{-}1}|^2 + |M_{j{-}1,k}|^2 \right)
\nonumber \\
A_{jk}^{(4)} &=& 2 \sqrt{jk} \, \omega' b\,m^2 
\left[ M_{j-1,k} M_{j,k-1} + M_{j,k-1} M_{j{-}1,k} \right]
\nonumber
\eeq 
Here, the electron Landau energy levels are given by
$E_n = \sqrt{1 + 2 n b},$
in units of \( m c^2 \). The longitudinal momentum of the pair is
$p = \sqrt{ \omega'^2 - (E_j + E_k)^2 }.$

The matrix elements \( M(j,k) \) account for the overlap of initial and final states and are defined as
\beq
M_{j,k} = (-1)^{j} \, \sqrt{ \frac{\underline{l}!}{\overline{l}!} } \,
\left( \frac{p_\perp^2}{2b} \right)^{|j-k|/2}
L_{\underline{l}}^{|j-k|} \left( \frac{p_\perp^2}{2b} \right)
\eeq
where 
$$\overline{l}=\max(j,k),\quad
\underline{l}=\min(j,k),$$
\( L_n^m(x) \) are the generalized Laguerre polynomials and \( p_\perp = \omega' \sin \theta \) is the transverse momentum of the pair.

In practice, the sums over \( j \) and \( k \) converge rapidly and are truncated when the matrix elements become negligible. These expressions are used to compute the attenuation coefficients as functions of photon energy, angle, and field strength, which in turn are integrated to obtain escape probabilities and effective opacities.

\bsp 
\label{lastpage}
\end{document}